\documentclass[12pt]{spieman}  
\usepackage{amsmath,amsfonts,amssymb}
\usepackage{graphicx}
\usepackage{setspace}
\usepackage{tocloft}
\usepackage{lineno}
\usepackage{subcaption}
\usepackage{placeins}
\title{On-sky demonstration of an ultra-fast intensity interferometry instrument utilizing hybrid single photon counting detectors}
\author[a, *]{Verena G. Leopold}
\author[a]{Sebastian Karl}
\author[b]{Jean-Pierre Rivet}
\author[a]{Joachim von Zanthier}
\affil[a]{Friedrich-Alexander-Universität Erlangen-Nürnberg (FAU), Quantum Optics and Quantum Information, Staudtstr. 1, 91058 Erlangen, Germany}
\affil[b]{Université Côte d'Azur, Observatoire de la Côte d'Azur, CNRS, Laboratoire Lagrange, France}

\cftpagenumbersoff{figure}
\cftpagenumbersoff{table} 
\begin{document} 
\maketitle

\begin{abstract}
Intensity interferometry is a reemerging astronomical technique for performing high angular resolution studies at visible wavelengths, benefiting immensely from the recent improvements in (single) photon detection instrumentation. Contrary to direct imaging or amplitude interferometry, intensity interferometry correlates light intensities rather than light amplitudes, circumventing atmospheric seeing limitations at the cost of reduced sensitivity.
We developed an ultra-fast, single photon counting and highly stable intensity interferometry instrument for 1\,m class optical telescopes. The instrument records on sky the expected stellar photon rates and reaches the temporal coherence times as measured in the laboratory. In addition, all components, especially the photon detection hardware, of the instrument are easily upgradeable with custom hardware currently being developed.
The collimated telescope output is spectrally filtered via an ultra narrow band pass of 2\,nm at a central wavelength of 405\,nm. We use hybrid photon detectors (HPDs) for single photon detection and a constant fraction discriminator (CFD) for signal conditioning. A time to digital converter (TDC) is used for time stamping. The combination of HPDs and CFDs is optimized for large active area and high timing resolution.
We successfully measured photon bunching of three bright A-type stars - Vega, Altair and Deneb at the $1.04\,$m Omicron telescope of C2PU (Centre Pédagogique Planète Univers) at the Calern Observatory in the south of France. In all cases the observed coherence time fits well to both the pre-calculated expectations as well as the values measured in preceding laboratory tests. We obtained the previously estimated photon count rates at the telescope and achieved highly stable coupling of the star light to the detectors. 
Utilizing a new class of large area single photon detectors based on multichannel plate amplification, high resolution spatial intensity interferometry experiments are within reach at $1\,$m diameter class telescopes within one night of observation time for bright stars.
\end{abstract}

\keywords{instrumentation:  interferometers -- methods: statistical -- site testing -- techniques: interferometric -- stars: fundamental parameters}

{\noindent \footnotesize\textbf{*}
Verena G. Leopold,  \linkable{vero.leopold@fau.de} }

 

\section{Introduction}
The history of stellar intensity interferometry starts with the growing interest in radar technology during the second world war. During that time A.J.F. Siegert looked at the fluctuations in signals finding a relation between the spatio-temporal correlations of the electric fields and of the intensities, today often called the Siegert relation, widely used in intensity interferometry \cite{siegert1943fluctuations}. Using this relation, 
Hanbury Brown and Twiss proposed a new type of interferometer to determine the diameter of thermal sources such as stars \cite{brown1954lxxiv} by measuring the correlation between photons in two separate beams of light \cite{brown1956correlation}.
Shortly afterwards they were able to estimate the angular diameter of Sirius A to 
$0.0063\,''$, a unique precision at that time \cite{brown20132}. Later, the first stellar intensity interferometer at the Narrabri Observatory in New South Wales used the described technique in order to measure the angular diameters of 15 stars \cite{brown1967stellar, hanbury1967stellar}.\\\newline
Recently, stellar intensity interferometry established itself as a complementary technique of current and planed imaging atmospheric Cherenkov telescopes (IACTs) as operation in the presence of bright moonlight is detrimental to observe Cherenkov light but harmless to intensity interferometry measurements \cite{le2006optical}. Prominent examples are VERITAS, MAGIC, ASTRI mini array and H.E.S.S. as well as the planned CTA. 
VERITAS was able to determine the angular diameter of $\beta$ Canis Majoris and $\epsilon$ Orionis with a precision of better than 5\% \cite{abeysekara2020demonstration}.
The VERITAS Stellar Intensity Interferometer (VSII) was further improved to increase its sensitivity \cite{kieda2021status} and since the beginning of 2019 continuing science observations and surveys of stellar targets are carried out
\cite{kieda2021veritas}. 
For the IACTs of Major Atmospheric Gamma-Ray Imaging Cherenkov (MAGIC) an optical setup is placed on top of the cameras of the two telescopes \cite{acciari2020intensity}. This way MAGIC was able to measure photon correlations \cite{acciari2020optical} to identify the stellar diameters of three different stars: Adhara ($\epsilon$ CMa), Benetnasch ($\eta$ UMa) and Mirzam ($\beta$ CMa) \cite{delgado2021intensity}. After hardware modifications MAGIC transitions within seconds to optical interferometry measurements and has determined 22 stellar diameters \cite{abe2024performance}. In a similar manner, two telescopes of H.E.S.S were able to record spatio-temporal correlations of three different southern sky stars \cite{zmija2024first}.
The ASTRI mini array with its 9 telescopes and baselines between 100 and 700\,m is planned to be equipped with fast single photon counting instruments \cite{zampieri2022stellar}.
\\
Single photon counting detectors have also been used with small optical telescopes to perform intensity interferometry measurements \cite{10.1093/rasti/rzae002}.
An example for spatial photon correlation measurements at long baselines is the Asiago Observatory (Italy) with a projected baseline of $\approx 2$ km \cite{zampieri2021stellar}.
With the help of a fixed $1.5\,$m telescope
and a $1.0\,$m portable telescope a group from the Université Côte d’Azur was able to observe the extended environment of $\gamma-$Cas at the Calern Observatory in the south of France \cite{matthews2023intensity}.
Elliot Horch and his group used two out of three portable 0.6\,m Dobsonian telescopes equipped with single-photon avalanche diode detectors to observe a correlation peak at the level of $6.76\,\sigma$ \cite{horch2022observations}. 
\\\newline
In this paper we present our results of temporal photon correlations of A-type stars in the blue using hybrid single photon counting detectors (HPDs), and outline the feasibility of using the same setup and telescope site for recording spatial correlations with two telescopes. We first describe the theory used to perform the stellar intensity correlations in Sec.\,\ref{sec:theory}. The telescope site and its specifications will be presented in Sec.\,\ref{sec:telescope} followed by a detailed explanation of the setup in Sec.\,\ref{sec:setup}. Details on the $2f-2f$ configuration in order to test the setup in the laboratory are laid out in Sec.\,\ref{sec:2f2f} and the corresponding results in Sec.\,\ref{sec:lab_results}. Our measurements of photon bunching for three different A-type stars can be found in Sec.\,\ref{sec:bunching}. In Sec.\,\ref{sec:advantages} we finally discuss the advantages of our setup and its feasibility of performing spatial correlations.
\section{Theory and experimental setup}
In the following we will briefly explain the theory behind stellar intensity interferometry that is needed for the evaluation of our observed temporal photon correlations. The specifications and properties of the telescope site will be outlined. We introduce our experimental setup and sketch our tests in the lab.
\subsection{Stellar Intensity Interferometry}\label{sec:theory}
The keyplayer in stellar intensity interferometry is the second order coherence function. It is defined as
\begin{equation}\label{eq:siegert}
\gamma^{(2)}(\tau)=\frac{\langle I(t) I(t + \tau)\rangle}{\langle I(t)\rangle^2}=1+|\gamma^{(1)}(\tau)|^2,
\end{equation}
where for the nominator and denominator averages are taken over time \cite{labeyrie2006introduction}. $I(t)$ denotes the time dependent intensity and $\gamma^{(1)}(\tau)$ is the first order coherence function also referred to as the complex degree of coherence which can be directly measured in amplitude interferometry. The second part of Eq.~\ref{eq:siegert} is called the Siegert relation that holds for thermal light sources such as stars and connects the second order coherence function to the modulus of the complex degree of coherence \cite{siegert1943fluctuations}. A direct consequence of this relation for thermal light is the effect of "photon bunching". Simply speaking, it is more likely to detect a second photon at a detector when a first photon was already recorded within the light coherence time.
For example, for a star $\gamma^{(2)}(\tau\rightarrow\infty)=1$ and therefore $\gamma^{(2)}(\tau=0)>\gamma^{(2)}(\tau\rightarrow\infty)$ holds with $\tau$ the delay in time. The width of this intensity correlation excess around zero time delay is proportional to the star's coherence time $\tau_c$. Moreover, the Wiener-Khinchin theorem relates the electric field spectrum $S(\omega)$ of a time-dependent wave to the Fourier transformation of its temporal autocorrelation function \cite{labeyrie2006introduction}. Hence, we can express the first order coherence function $\gamma^{(1)}(\tau)$ as follows
\begin{equation}
\gamma^{(1)}(\tau)=\int S(\omega)e^{i\omega\tau}\,d\omega.
\label{eq:wiener}
\end{equation}
The Siegert relation together with the Wiener-Khinchin theorem is a powerful tool in stellar intensity interferometry.
We choose the coherence time $\tau_c$ to be the integral of the obtained temporal correlation peak
\begin{equation}
    \tau_c = \int_{-\infty}^\infty |\gamma^{(1)}(\tau)|^2\,d{\tau}.     
    \label{eq:coh_time_def}
\end{equation}
This definition enables to measure the coherence time by integration even if the timing resolution of the photon detection system $\Delta t$ is orders of magnitude larger than $\tau_c$. In this case the measurement differs widely from the expectation in Eq.~\ref{eq:siegert}  and Eq.~\ref{eq:wiener}. In order to calculate the expected measurement result $\gamma^{(2)}_\mathrm{meas}$, the second order coherence function has to be convolved with the detection system timing uncertainty $P_\mathrm{diff}(t)$: 
\begin{equation}
\gamma^{(2)}_\mathrm{meas}(\tau) = \left(\gamma^{(2)}(t) \ast P_\mathrm{diff}(t)\right)(\tau).
\label{eq:meas_res_estimation}
\end{equation}
This convolution leads to a peak with an amplitude $<1$ and a shape similar to the timing jitter. The area under the curve, and thus the coherence time, are however unchanged if  $P_\mathrm{diff}(t)$ is properly normalized. Note that, when considering only a single polarization mode, the height of the bunching peak is reduced to $\sim\tau_c/\Delta t$. Using the relation $\tau_c = \lambda^2/(c \times \Delta \lambda)$ \cite{mandel_wolf_book}, we can estimate the coherence time of any band pass larger than $\Delta \lambda=0.5\,$nm (FWHM) to $\tau_c \ll 1\,$ps. As typical detection systems have a timing jitter around ns to $100\,$ps, the detected signal amplitude will be drastically reduced. However, our detection system has a much lower timing jitter of $42.4\,$ps resulting in a better timing resolution, when compared with similar measurements.
\\
Resolving such a signal with diminished amplitude within a correlation histogram can be a major challenge. The minimal amount of observation time necessary to yield a certain signal to noise ratio (S/N) can be estimated under the following assumptions: 1) The observed count rates $n$ are equal for all detectors and constant over the whole measurement time $t_\mathrm{meas}$; 2) The bin error for a histogram bin with $N$ clicks is $\Delta N = \sqrt{N}$ (shot noise); 3) The measurement has a timing resolution $\Delta t$ and is binned into a correlation histogram with bin size $t_\mathrm{bin}$; 4) the shot noise error is equal for all bins comprising the bunching peak. Propagating the errors we can write the S/N as 
\begin{equation}
\mathrm{S/N} = \frac{\tau_c}{\Delta \tau_c} = \frac{\tau_c}{t_\mathrm{bin}\sqrt{\sum_{i=1}^{m} \left(\Delta \gamma^{(2)}(\tau_i)\right)^2} }.
\end{equation}
Here we used our integral definition of the coherence time as well as the Siegert relation from Eq.~\ref{eq:siegert} and propagated the shot noise error taking an index $i$ iterating over the $m$ bins that constitute the bunching peak. The detected coincidences $N$ per bin are equal to $N = n^2 \,t_\mathrm{meas}\, t_\mathrm{bin}$, where $n$ is the detected count rate at each detector. Assuming that the shot noise is equal for each of these bins we can simplify the expression 
\begin{equation}\label{eq:S/N}
\mathrm{S/N} = \tau_c \frac{1}{t_\mathrm{bin}\sqrt{m} } \sqrt{n^2 \,t_\mathrm{meas}\, t_\mathrm{bin}} = \tau_c \,n \sqrt{\frac{t_\mathrm{meas}}{\Delta t} }, 
\end{equation}
where we used the fact that the number of bins $m$ making up the correlation peak is equal to the ratio of the timing resolution $\Delta t$ and the bin size $t_\mathrm{bin}$ in the limit $\Delta t\gg\tau_c$. For this estimation to hold one has to assume the timing resolution to be twice the measured FWHM resolution to account for full sampling of the bunching peak. It should be pointed out that the count rate $n$ is proportional, whereas the coherence time $\tau_c$ is inversely proportional, to the optical bandwidth of the observed light field, hence the S/N is independent of the optical bandwidth to first order. Nevertheless, the usage of a narrow band pass can be beneficial to increase the height of the correlation peak above systematics inherent to the photon detection system. When substituting the conventional definition of the coherence time and spectral flux one arrives at the result of \cite{hbt_snr_1957}. \\
Finally, Eq.~\ref{eq:S/N} can be used to estimate the necessary measurement time for a given S/N. Assuming a coherence time of $0.2\,$ps (as expected by our setup), a count rate of $2\,$MHz, a detection resolution of  $40\,$ps and aiming for an S/N of 5, we estimate a measurement time of $t_\mathrm{meas} \approx 6.8\,$h.
\subsection{The Omicron telescope}\label{sec:telescope}
We performed measurements at the Omicron telescope of C2PU, part of the Observatoire de la Côte d'Azur at the plateau of Calern in the south of France. C2PU is located $06^\circ55'23''$\,East and $43^\circ45'13''$\,North at an altitude of $1270\,$m. C2PU offers twin telescopes with a primary mirror diameter of $1.04\,$m at a separation of $15\,$m. Our measurements were performed in the west dome of C2PU on the Omicron telescope. The telescope uses an equatorial yoke mount and has a Cassegrain secondary focus with a focal length of $13\,$m and a focal ratio of $F/12.5$.
\subsection{Experimental setup}\label{sec:setup}
\begin{figure}%
\centering
\includegraphics[alt={Drawing of the used setup with all optical components and detectors displayed.}, width=5in]{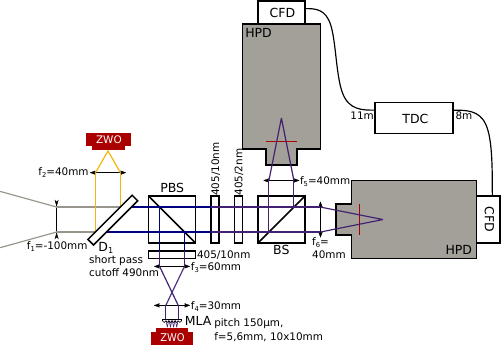}%
    \vspace{1ex}%
    \caption{Schematic of the used setup. The light enters from the left. The color split yellow light hits the ZWO ASI120MM-S guiding camera and the polarized light is used in the Shack-Hartmann arm in order to monitor the collimation of the wavefront. On the main axis the light passes through two ultra narrow band pass filters with a central wavelength of $405\,$nm before being split onto the two hybrid single photon counting detectors (HPDs). The signal output is fed through a constant fraction discriminator (CFD) and connected to the time to digital converter (TDC). Note that the signal cables have a length difference of $3\,$m.}
    \label{fig:setup}
\end{figure}

\begin{figure}
    \begin{subfigure}{0.5\textwidth}
	\centering
    \includegraphics[alt={Cuboid support structure with the opitcal setup including the detectors and their movement assembly inside.}, width=3.3in]{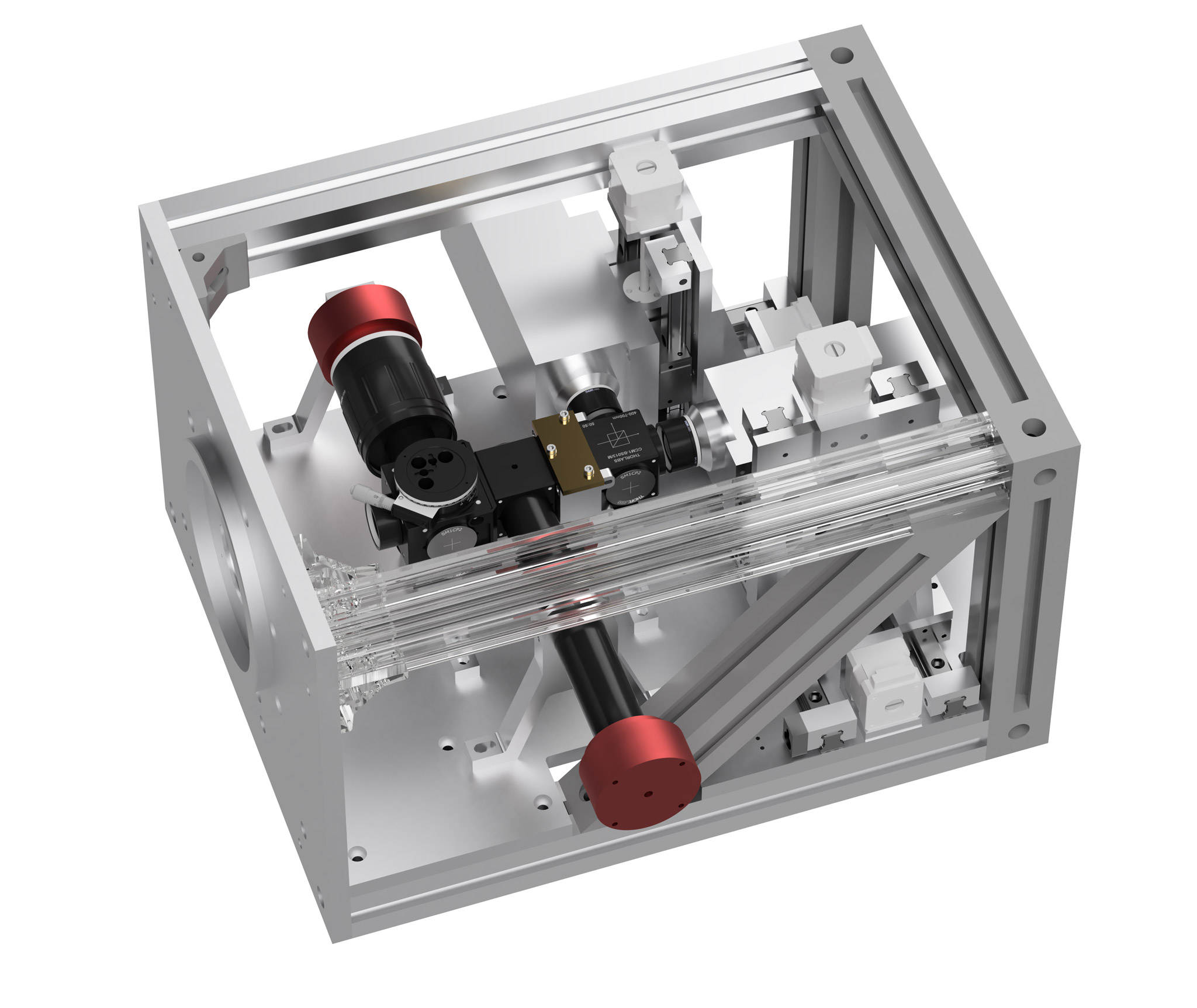}
    \caption{}
    \label{fig:setup_render}
\end{subfigure}%
\begin{subfigure}{0.5\textwidth}
	\centering%
    \includegraphics[alt={Picture of the setup attached to the telescope output.}, width=\textwidth]{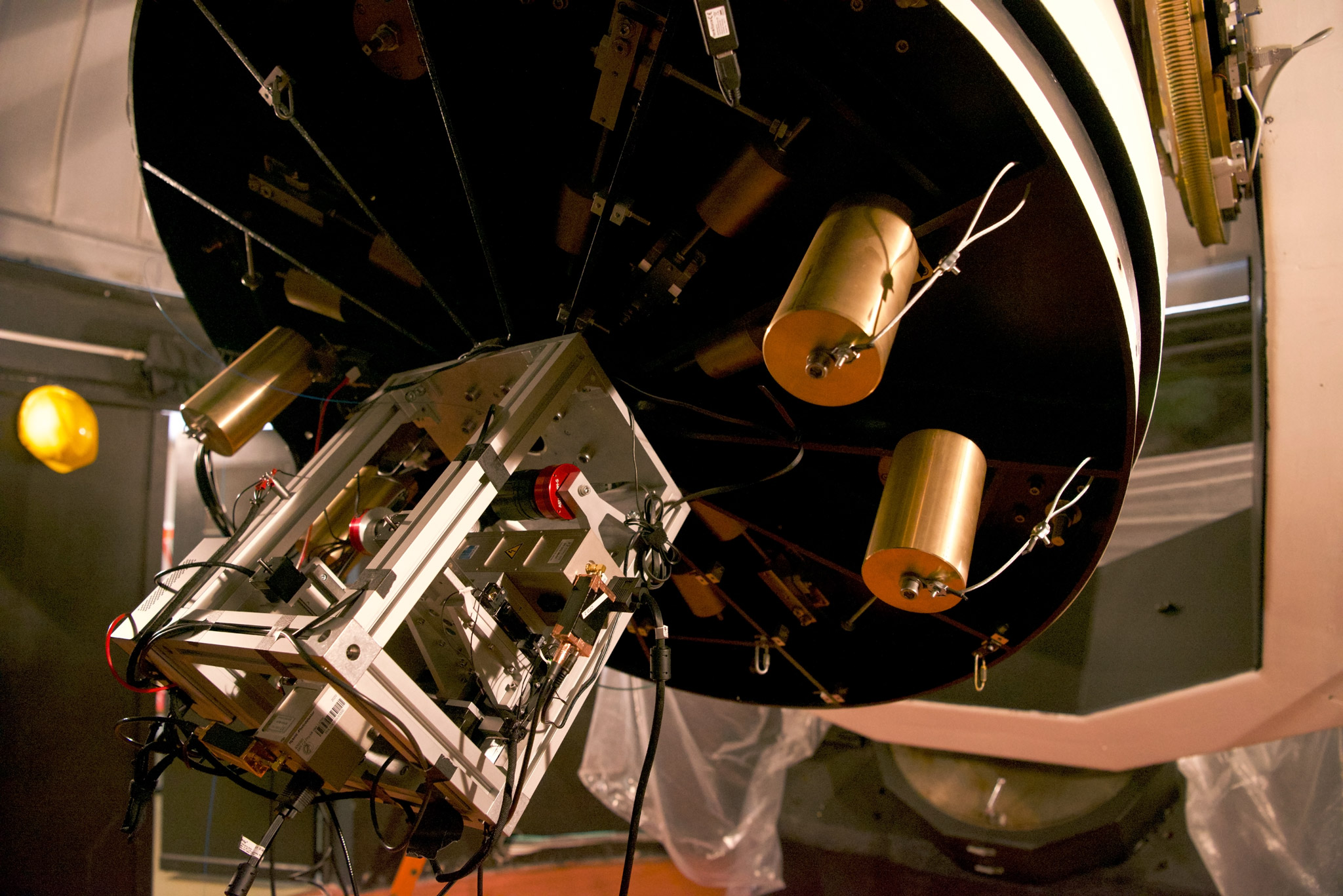}%
    \vspace{2.2ex}%
    \caption{}
    \label{fig:setup@cope}
\end{subfigure}%
    \vspace{1ex}%
\caption{(a) CAD render of the used setup and its support structure. The light enters through the hole on the left. The ring at the front plate is used to center the setup inside the telescope flange, see also Fig.~\ref{fig:setup@cope}. One of the aluminum profiles is transparent to enable a better view onto the setup. The optical components are mounted to the ground plate, see Fig.~\ref{fig:setup} for a detailed explanation of the optical setup. The hybrid single photon counting detectors (HPDs) are motorized via two stepper motors enabling movement in the xy-plane. (b) The setup is mounted via $10$ M12 screws to the telescope's output flange. The light leaves the telescope at the center of its bottom and is directly coupled into our setup hitting the first lens for collimation. The brass cylinders are the necessary counter weights in order to balance the telescope after mounting the approximately $27\,$kg setup.}
\end{figure}

The setup used in order to measure temporal correlations is shown in Fig.~\ref{fig:setup}. The light from the telescope enters the setup from the left and is collimated by an achromatic lens with a focus of $-100\,$mm resulting in a beam diameter of $8\,$mm. This is a major change compared to one of our previous setups used at an optical telescope that featured a deflection mirror at the telescope flange \cite{10.1093/rasti/rzae002}. A short pass dichroic mirror (D$_1$) with a cutoff wavelength of $490\,$nm transmits the light used for the intensity correlations. The reflected light is focused via an achromatic lens with a focus of $40\,$mm onto a ZWO ASI120MM-S camera (ZWO) used for guiding.  The transmitted light is passed through a polarizing beam splitter (PBS) which is useful in order to increase the contrast of the bunching peak. Besides it separates light for wavefront analysis without a discernible loss in measurement S/N. The light reflected by the PBS impinges on a self-made Shack-Hartmann sensor. A two lens system is used in order to reduce the beam diameter before a $10\times 10\,$mm microlens array (MLA) with a lens pitch of $150\,\mu$m and a focal length of $5.6\,$mm. A second ZWO is placed in the focal plane of the MLA monitoring the spot diagram, that is the collimation quality of the observed wavefront. This is an important information as an uncollimated beam would degrade the ultra narrow band pass. 
On the main axis the light is first pre-filtered by a Thorlabs band pass with a central wavelength of $405\,$nm and $10\,$nm bandwidth before passing through a $2\,$nm ultra narrow band pass with a central wavelength of $405\,$nm by Alluxa, both shown as rectangles in Fig.~\ref{fig:setup}. This combination is crucial to minimize the possible out-of band transmission. A $50:50$ non-polarizing beam splitter (BS) separates the light onto the two hybrid single photon counting detectors (HPDs) from Becker\&Hickl with a quantum efficiency of $22.7\,\%$, see measurement plotted in  Appendix \ref{sec:hpd_qe}. The signal output from the HPDs is fed through a custom Constant Fraction Discriminator (CFD) produced by Photonscore to generate NIM pulses. This custom CFD enables us to use long cables and a time to digital converter (TDC), the quTAG from quTools, that has a higher timing resolution compared to the time to amplitude converter from our previous work \cite{10.1093/rasti/rzae002}. Due to technical limitations the quTAG introduces a dead time of $40\,$ns much longer than the $5\,$ns dead time of the detectors. With this rather large dead time start-stop correlations are sufficient for histogramming the photon delay times. In order to account for the $6\,$mm active area of the HPDs, lenses with a focal length of $40\,$mm slowly converge the beam towards the photocathode, shown as red line inside the HPD. Hence, we are not focusing onto the photocathode but rather only gently decrease the beam diameter to fit the size of the cathode leading to a better beam stability. 
The CFD threshold is set to a similar value as suggested by the HPD manufacturer, whereas the zero crossing is optimized to obtain a high time resolution of around $30\,$ps. These two settings lead to dark counts on the order of $50\,$Hz. The NIM output signal of the CFD then travels to the quTAG through a $11\,$m and $8\,$m signal cable respectively to perform the correlations. The cable delay of $3\,$m is necessary for the signal to lie inside a nearly shot-noise limited environment, see Appendix \ref{sec:lab_test_results}. Post-processing is done via a PC.\\
\newline
A CAD render of the setup and the necessary support structure is shown in Fig.~\ref{fig:setup_render}. The support structure is mounted directly to the output telescope flange, see Fig.~\ref{fig:setup@cope}. The light from the telescope directly enters the setup having a beam diameter of $8\,$mm when passing through the first lens. The HPDs are movable in the xy-plane, using two stepper motors per detector. Note that due to the high stability of the setup the detector positions once aligned and fixed in the lab, stayed in this position throughout the entire measurement. The setup is designed to fit the constraints of the telescope, these are. weight and dimensions. Counter weights have been used to balance the telescope again after mounting the setup with a weight of approximately $27\,$kg.%
\subsection{$2f-2f$ configuration test}\label{sec:2f2f}
\begin{figure}
	\centering
    \includegraphics[alt={Schematic of configuration with an inset of the pupil diaphragm. The pupil diaphragm looks like a circular mask with a smaller circle inside. The distance between the source (end point of the multi mode) and the lens is 400 mm.}, width=5in]{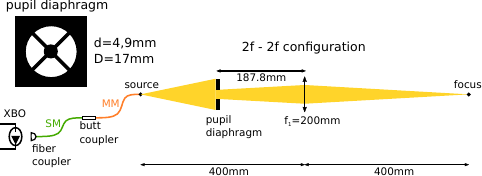}%
    \vspace{1ex}%
    \caption{Schematic of the $2f-2f$ configuration and the used pupil diaphragm. Light from the xenon short-arc lamp (XBO) is coupled into a single mode fiber (SM) and then into a multi mode fiber (MM) before diverging until the telescope simulator (pupil diaphragm). The resulting mode is similar to the beam profile expected at the Omicron telescope. A lens with a focal length of $200\,$mm focuses the light into the experimental setup. In order to tune the count rate, ND-filters can be placed directly after the Source.}
    \label{fig:2f2setup}
\end{figure}
Before on-sky observations, the setup was pre-aligned and tested in the lab. An Osram XBO $75$\,W$/2$ OFR, that is a xenon short-arc lamp (XBO), was used as an artificial star. In order to obtain a similar beam profile as expected at the telescope a $2f-2f$ configuration was used, see Fig.~\ref{fig:2f2setup}. The light from the XBO is first coupled into a single mode fiber in order to collect only a single spatial mode. This single mode fiber is coupled directly to a short piece of $50\,\mu$m mode field diameter step index multimode fiber. The tip of the multimode fiber is used as source from which the light diverges freely. Its diameter is chosen such that it emulates a $25'$ field of view, while the short length suppresses the occurrence of adverse timing effects caused by modal dispersion. The diverging beam hits a 3D-printed pupil diaphragm resembling the constraints at the Omicron telescope. The pupil diaphragm, also referred to as the telescope simulator, has a cut out circle with a diameter of $D=17\,$mm and a blocking inner circle with a diameter of $4.9\,$mm to simulate the secondary mirror, see inset of Fig.~\ref{fig:2f2setup}. The resulting mode is focused via a lens with a focal length of $200\,$mm. The experimental setup is placed around the focus obtaining a collimated beam after its first lens. Note that all components of the $2f-2f$ configuration were fixed to a $15\,$mm range micrometer stage. This has the advantage of tuning the focus, hence, facilitating the alignment. The count rate at the HPDs was adjusted via ND-filters that were placed directly after the source. 
\section{Results and discussion}
\subsection{Lab test}\label{sec:lab_results}
Before testing our setup, we measured the timing resolution of our detection system (TDC, HPDs and CFDs) in the laboratory using a pulsed fs-laser to be $42.4\,$ps (see Appendix ~\ref{sec:timing jitter}), which is low compared to typical detection systems that are on the order of 100ps to ns. Together with the filter transmission spectra supplied by the filter manufacturers this measurement allows the calculation of both the expected coherence time and expected shape of the bunching peak using equations Eq~\ref{eq:siegert}, \ref{eq:wiener}, \ref{eq:coh_time_def}, and \ref{eq:meas_res_estimation}. We calculate the expected coherence time to $0.212\,$ps, and plot the expected shape of the bunching peak in each of the measurement results shown below. A $2f-2f$ configuration test of our setup, as described in Sec.~\ref{sec:2f2f}, confirmed our expectation and revealed a coherence time of $(0.21 \pm 0.02)\,$ps (see Fig.~\ref{fig:lab_test}).
\subsection{Bunching of A-type stars}\label{sec:bunching}
\begin{figure}
    \begin{subfigure}{0.5\textwidth} 
	\centering
    \includegraphics[alt={Scatter plot of the measured bunching signal in units of the second order coherence function minus 1 over time in ps for Vega. A peak is visible at zero ps.}, width=3.3in]{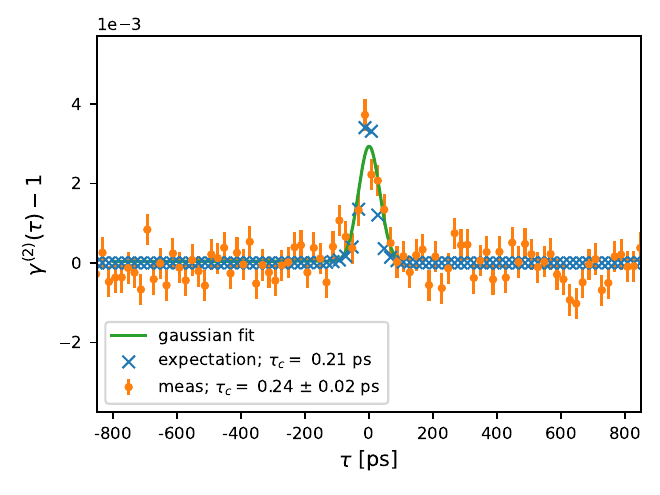}
    \caption{Vega}
\end{subfigure}
\begin{subfigure}{0.5\textwidth}
	\centering
    \includegraphics[alt={Scatter plot of the measured bunching signal  in units of the second order coherence function minus 1 over time in ps for Altair. A peak is visible at zero ps. The measurement looks more noisy compared to the plot obtained for Vega.}, width=3.3in]{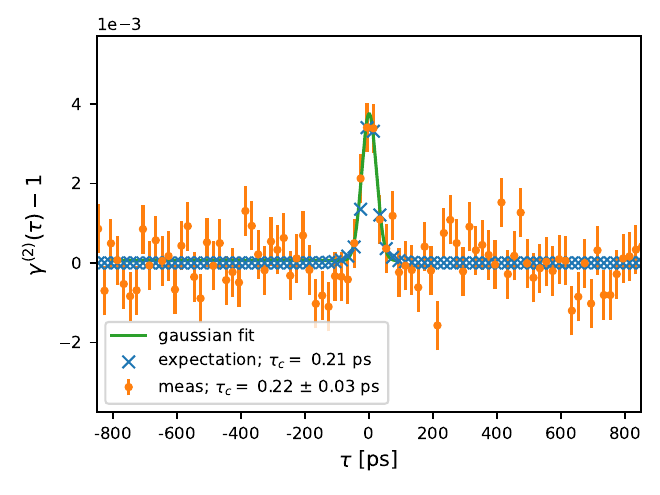}
    \caption{Altair}
\end{subfigure}\\
\begin{subfigure}{\textwidth}
\centering
    \includegraphics[alt={Scatter plot of the measured bunching signal  in units of the second order coherence function minus 1 over time in ps for Deneb. A peak is slightly visible at zero ps inside a noisy background.}, width=3.3in]{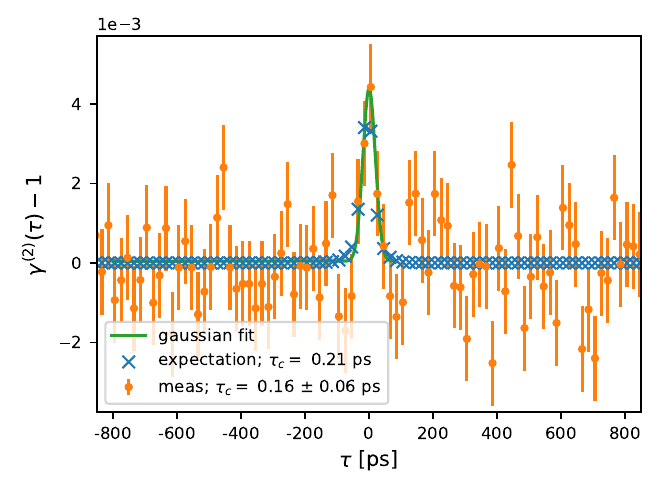}
    \caption{Deneb}
\end{subfigure}%
    \vspace{1ex}%
\caption{Bunching signal measured for different stars shown in orange with corresponding error bars. The expected measurement result is shown as blue crosses. A Gaussian fit is used for centering purposes (green line). The measurement result fits well to the expected coherence time of $0.21\,$ps.}
\label{figure:stars}
\end{figure}

\begin{figure}
    \begin{subfigure}{0.5\textwidth}
	\centering
    \includegraphics[alt={Root mean square error (RMSE) (top) and re-scaled residuals of the RMSE from shot noise (bottom) plotted over the number of counts per bin N. Top plot shows a square root dependence for the expectation as well as the result for Vega. The curves start to differ for higher N. Bottom plot depicts the growing residuals for all three stars.}, width=\textwidth]{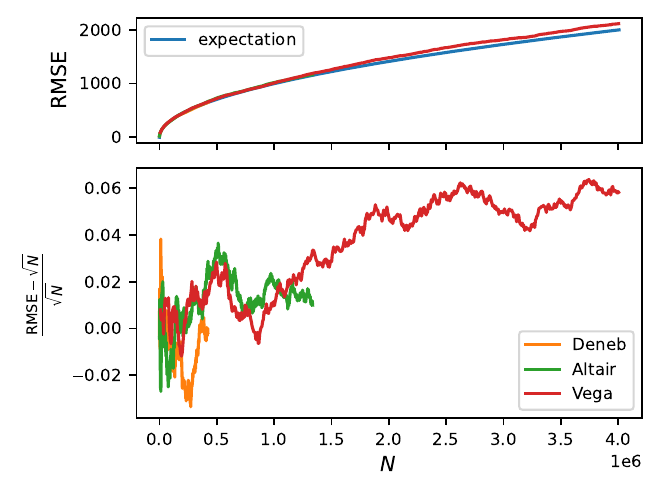}
    \caption{}
    \label{fig:rsme}
\end{subfigure}
\begin{subfigure}{0.5\textwidth}
    \centering
    \includegraphics[alt={Count rate at the detectors over time for one measurement night. Stars were observed in the following order: Vega, Altair, Deneb. Most of the time was spend on Altair. Continuous count rate curves visible for each star.}, width=\textwidth]{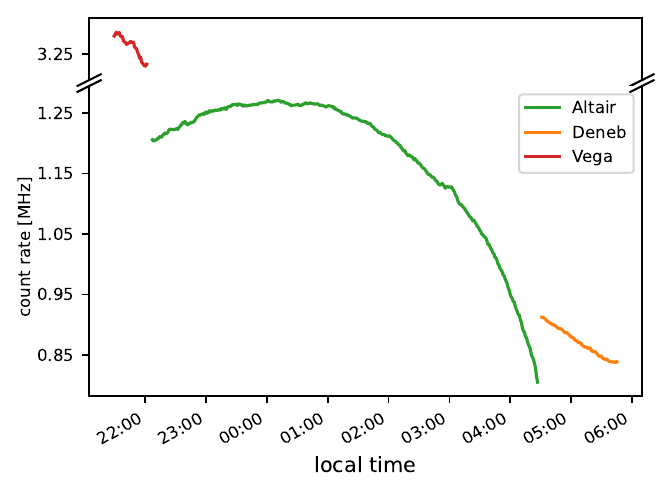}
    \caption{}
    \label{fig:countrate}
\end{subfigure}%
    \vspace{1ex}%
\caption{(a) Root mean square error (RMSE) (top) and re-scaled residuals of the RMSE from shot noise (bottom) plotted over the number of counts per bin $N$. The RMSE is expected to be shot noise $\mathrm{RMSE} = \sqrt{N}$ for $N$ counts per histogram bin. The upper plot shows deviations from shot noise in Vega's curve for higher count rates. This is due to the limits of the detection system. These deviations are shown in greater detail for all observed stars in the lower plot. While the residuals are initially close to zero for all stars, the relative residuals are larger for targets with higher detected count rates. Even for the high S/N measurement recorded for Vega, the RMSE is at most 6\% larger than the shot noise expectation. (b) Count rate at the detectors over time during the night of the 13th of August. The count rate is averaged every $30\,$s. Three different stars were observed. One can clearly see that each measurement was quite stable, meaning. tracking of the star was never lost. This is due to the stability of the equatorial yoke mount of the telescope and the rather large active area of the HPDs. The time-evolutions of the count rates for Altair and Deneb are
mainly driven by the air mass evolution due to sidereal motion. For Vega, the flux variation is more likely to be due to atmospheric transmission fluctuations.}	
\end{figure}

Once the setup was mounted on the Omicron telescope, we observed 
three bright A-type stars: Vega, Altair, and Deneb. The star light was measured starting the night of 9 August 2023 and ending the night of 18 August 2023. During one night no measurements could be taken (16 August 2023) due to bad weather conditions, and during one night the results were compromised by a malfunction of the quTAG (11 August 2023), leaving us with eight good measurement nights. Throughout all these nights the sky was very clear, with only some slight high layer clouds in rare instances. A summary of the measurements is given in Table~\ref{tab:star_overview}, showing the apparent magnitudes, the coherence time as well as its error, the observation time, and S/N for each star. A significant portion of our observation time ($12.1\,$h) was spent on Vega, aiming for bunching at a high S/N. In addition Vega was observed at the beginning of each night starting 2023-08-12 for roughly 30 minutes, except for 2023-08-18, using the large photon count rate to ensure our detection electronics were working correctly. The largest part of our observation time was spent on Altair ($29.9\,$h), attempting to resolve bunching for a star with slightly larger apparent magnitude. We spent the final nights of our campaign observing Deneb, in order to observe bunching for an even dimmer star. \\
The bunching measurement results for each of the stars are plotted in Fig~\ref{figure:stars}. These plots show the measurement result with error bars generated by measuring the root mean square error (RMSE) of the baseline of the correlation histogram far from the correlation peak. We further show the expected measurement result (blue crosses), as well as a Gaussian fit to the data which is only used for centering the bunching peak with respect to zero time delay. In all cases  the measurement results fit to our expectation very well, deviating from the expectation by less than $1\,\sigma$. The coherence times listed in Table~\ref{tab:star_overview} and shown in the plot legends were obtained by numerical integration between the $3\,\sigma$ borders of the expected measurement result. This guarantees equal integration borders for all three measurements. The uncertainties of these coherence times were obtained via appropriate error propagation of the bin uncertainties shown as error bars. Due to the differing count rates owed to different apparent magnitudes and the varying observation times each bunching measurement shows a different S/N. We obtained the highest S/N for Vega of 12, and managed to observe bunching with an S/N of 7.3 and 2.7 for Altair and Deneb respectively. \\
Table~\ref{tab:countrate} lists the median, minimum, and maximum count rates of the observed stars evaluated over the whole observation range without any further filtering, as well as the standard deviation of the measured count rates.  
We need to estimate the expected flux at the telescope and through our optical setup and compare it to the obtained count rates in order to classify the quality of our measurements. Since we are close to the transition between the U and B band of astronomical wavelengths, we average the flux within these two bands to $1074\,\mathrm{photons} \, \mathrm{s}^{-1}\,\text{\r{A}}^{-1}\,\mathrm{cm}^{-2}$ \footnote{Photon fluxes taken from \url{https://www.astronomy.ohio-state.edu/martini.10/usefuldata.html}}. Taking into account the primary mirror area of the C2PU Omicron telescope and our filter bandwidth the incident photon rate through our band pass is approximately $183\,$MHz. Accounting for losses at the telescope mirrors, the secondary mirror obstruction, losses within the optical elements as specified by the manufacturer, our detectors quantum efficiency, and the event loss by the detector dead time we should observe a count rate of $3.96$\,MHz per detector for Vega. Since a median count rate of $3.12\,$MHz was observed, that is 79 percent of the expectation, we conclude that with our setup an effective coupling of the starlight to our detectors was obtained. The count rates observed for both Altair and Deneb are within the expectations when comparing their apparent magnitude in the B and U band to Vega. \\
Despite introducing an appropriate cable delay and taking appropriate measures to shield our electronics from the RF background of the telescope dome, we observed some deviation from shot noise when evaluating the RMSE for each of our measurements. The observed RMSE over the number of counts per bin $N$ is shown in Fig.~\ref{fig:rsme}. Ideally all RMSE curves should show shot noise closely following a square root curve. For Vega this curve is displayed in the top part of Fig.~\ref{fig:rsme}. Plotting the residuals divided by their shot noise value in the lower part of Fig.~\ref{fig:rsme} shows that the relative residuals are initially close to zero for all observed stars, but around $N = 0.25 \times 10^6$ the deviations grow. Furthermore, the relative residuals are larger for targets at higher count rates, with Vega deviating 6\% from shot noise at the end of its observation time, Altair deviating 1.5\%, and Deneb nearly not deviating at all. While this discourages observations at higher detected photon rates with our current time tagger, we also note that the relative residuals seem to be rather stable once settled in, still allowing high S/N bunching measurements. Similar deviations from shot noise have been measured in our laboratory test (cf. Appendix \ref{sec:lab_test_results}). We thus conclude that this deviation from shot noise is most likely intrinsic to our detection electronics.

\begin{table}
	\centering
	\caption{Overview of the observed A-type stars and their properties. Apparent magnitudes were taken from \protect\cite{bohlin2004hubble} and \protect\cite{ducati2002vizier}.}%
    \vspace{1ex}%
	\begin{tabular}{lcccc} %
		\hline \hline
		Star & App. Magnitude & meas. $\tau_c$ [ps] & Obs. Time [h]& S/N\\
		\hline
		Vega & 0.026 & $0.24\pm 0.02$ & 12.1 & 12\\
		Altair & 0.76 & $0.22\pm 0.03$ & 29.9 & 7.3\\
		Deneb & 1.25 & $0.16\pm 0.06$ & 17.3 & 2.7\\
		\hline
	\end{tabular}
 \label{tab:star_overview}
\end{table}
\subsection{Advantages of our setup}\label{sec:advantages}
The short optical path of our setup, that is the distance from the entrance point of the light to the detector allows a straightforward alignment. This is a major improvement compared to one of our previous setups that featured a deflection mirror at the telescope flange \cite{10.1093/rasti/rzae002}. The alignment is done by adjusting merely the telescope's focus and its pointing/tracking of the star, in order for the beam to impinge centrally. All optical components were pre-aligned in the lab and no further adjustments of optics inside the setup were necessary.\\
The use of HPDs produced by Becker\&Hickl has the advantage of observing stars in the B-band. This can contribute to new science as observations in the blue, here at $405\,$nm, are rather rare and can be used to test or verify commonly used models for stars in the blue \cite{sackrider2022stellar} \cite{acharyya2024angular}.\\
Another benefit is the high stability of the measurement during the observation. A typical night is shown in Fig.~\ref{fig:countrate}. One can see that tracking of the observed star was never lost. This is due to the rigid mount of the telescope and to the large active area of the HPDs compared to avalanche photo diodes (APDs). This facilitates tracking during the night; one can even see the typical slope of the count rate for a star moving along the sky. The altitude of Altair decreases during the night and therefore, the telescope observes closer to the horizon, meaning the light will be attenuated by additional atmospheric layers, visible as a decrease in count rate. Note that due to mechanical restrictions of the Omicron telescope as well as the location of the stars on the sky, Deneb and Altair could not be observed the whole night. The large area of our detectors also decreases the susceptibility of our setup to astronomical seeing. Even in adverse seeing conditions of $\sim 3\,''-\,\,4\,''$, our setup would be able to couple starlight to the detectors with the same efficiency as in good seeing conditions, as our field of view is $2\,'$ and the beam on the detectors only needs to be stable within $1\,$mm. \\
Table~\ref{tab:countrate} shows the observed count rates for all stars evaluated over all measurement nights. The median, maximum, and minimum count rate are calculated in a standard fashion, taking all data into account uncorrected. Note that the rather small minima for the count rates are due to bad weather conditions, these are clouds covering the star for a short time during the night. The deviation in Table~\ref{tab:countrate} is calculated after correcting for the deterministic change in count rate due to the star's motion along the night sky. For all observed stars, the standard deviation of the count rate is orders of magnitude smaller than the median observed count rate, indicating fairly stable coupling of the star light to our detectors. For Vega and Deneb, we observe a larger fluctuation in count rate than for Altair, indicating slightly worse observation conditions. This is also reflected by the more pronounced loss in count rate when comparing minimum to median count rate for Vega and Deneb with respect to Altair. 
\begin{table}[h!]
	\centering
	\caption{Overview of the observed A-type stars and their obtained count rates evaluated over all measuring nights. The median and maximum values are obtained by averaging over all count rates observed without any post-processing. To obtain a value for the standard deviation that represents the actual count rate fluctuations observed we first subtract a smoothing fit from the observed count rates before calculating the standard deviation. This is done in order to compensate for the deterministic change in count rate due to the star's movement along the night sky.}%
    \vspace{1ex}%
	\label{tab:countrate}
	\begin{tabular}{lcccc} %
		\hline
        \hline
		Star & Median & Max. & Min. &  Dev.\\
		\hline
		Vega &  $3.12\times 10^6$ & $3.56 \times 10^6$ & $0.16 \times 10^6$ & $0.17 \times 10^6$ \\
		Altair & $1.14\times 10^6$ & $1.27 \times 10^6$ & $0.68 \times 10^6 $& $0.0035 \times 10^6$ \\
		Deneb & $8.19\times 10^5$ & $10.19 \times 10^5$ & $0.95 \times 10^5$ & $0.58 \times 10^5$ \\
		\hline
	\end{tabular}
\end{table}
\section{Conclusion and outlook}
We have successfully developed an ultra-fast intensity interferometry instrument using off the shelf components with a high timing resolution of $42.4\,$ps. We measured temporal photon correlations and photon bunching of three different bright A-type stars - Vega, Altair, and Deneb. In all cases the observed coherence time fits well to both the pre-calculated expectations as well as the value measured in preceding laboratory tests. The maximum S/N was 12, achieved for Vega after $12.1\,$h of observation. For Altair and Deneb, the S/N was significantly lower, that is 7.3 and 2.7, respectively. Taking into account the apparent magnitudes of the observed stars in the U and B bands, we obtained more than 79 percent of the expectation for each target due to the efficient coupling of the starlight to our detector. \\
In that respect, our setup shows notable advantages with respect to fiber coupled detector configurations. These will also hold when measuring spatial photon correlations, that is when two setups are used at two telescopes. Due to the large active area of our photodetectors the entire setup can be easily aligned at the telescope without need for readjustments of the optical components. The large area also allows for stable coupling of the light field to the detectors, decreasing the sensitivity to both tracking errors and atmospheric seeing resulting in a nearly autonomous operation at the telescope. Spatial correlation measurements are feasible on the instrumentation side and promising for testing common astronomical models for stars in the blue. Our setup worked well at the Omicron telescope, hence, we plan to perform spatial correlation measurements in the near future in the blue at C2PU.\\
For these measurements, we plan to use a new type of detector based on a Photonis Hi-QE photocathode and multichannel plate (MCP) amplification. These detectors combine the advantages of a large active area of $8\,$mm, a good timing resolution of $<50\,$ps, and a nearly deadtime free operation. The upgrade to these detectors would increase our S/N per unit measurement time by a factor of $\approx 3$ due to the Hi-QE photocathode's enhanced quantum efficiency. The latter is 50 percent greater compared to conventional Bialkali photocathodes, such as the ones used in our HPDs. A further benefit is offered by the fact that such detectors are virtually deadtime free due to the ability of the large MCP to support multiple confined electron avalanches within a short timescale. This prevents the loss of potentially detected photons due to a previous detection, which is an advantageous feature even at the low event rates considered in this paper.\\ 
A promising avenue for spatial photon correlation measurements is to duplicate our optical setup, replace the HPDs with Hi-QE based detectors and use a self-developed TDC. When performing a measurement for more than $10\,$h using the quTAG, we saw that the re-scaled residuals increase compared to measurements at the telescope, see Appendix \ref{sec:lab_test_results} and Fig.~\ref{fig:RSME_residuals_lab}. This increase is believed to be due to the read-out electronics. As for measurements at the telescope we additionally saw that the residuals increase for higher count rates (Fig.~\ref{fig:rsme}). We are currently developing our own TDC aiming for a higher resolution at higher count rates. This TDC is dead time free (no additional $40\,$ns) and therefore does not loose photon detection events so that the histogramming can be done via multi-start multi-stop. The new detection system would thus have an even lower timing jitter compared to the shown measurements.\\
Using Hi-QE instead of HPD detectors and observing a magnitude $0$ A-type star such as Vega, spatial correlations with S/N 5 (10) are expected to be carried out within $0.9\,$h ($3.5\,$h) per baseline. This means that for a mag. $0$ A-type star and two $1\,$m diameter telescopes, high resolution intensity interferometry measurements within one observation night are within reach. For a slightly dimmer star such as Altair, the required observation time for the same S/N values increase to $6.5\,$h per baseline for an S/N of 5, which is still feasible given the combination of multiple observation nights. Measuring with two detectors per telescope will give access to four independent spatial correlation measurements with the same baseline, which can be added up to decrease the bin error by a factor of two over a single temporal correlation measurement. This will however not increase the S/N of the spatial photon correlation measurement, as the squared visibility for most bright A-type stars is about $0.5$ for the $15\,$m on-ground baseline of C2PU, reducing the bunching peak height by that factor. In fact these two contributions cancel out, leading to the same S/N as in a temporal correlation measurement using a single telescope.

\appendix    
\section{Timing jitter} \label{sec:timing jitter}
\begin{figure}
    \begin{subfigure}{0.5\textwidth}
    \centering
    \includegraphics[alt={Timing jitter of detection system over time in ps. The measurement shows a peak and is fitted via a gaussian. }, width=\textwidth]{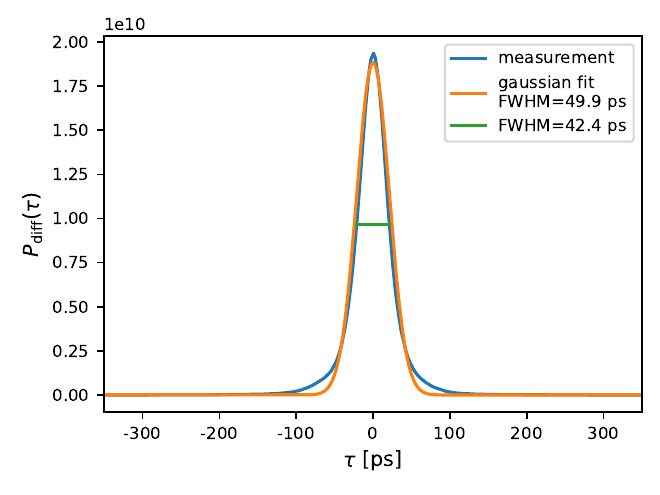}
    \caption{}
    \label{fig:timing_jitter}
    \end{subfigure}
    \begin{subfigure}{0.5\textwidth}
    \centering
    \includegraphics[alt={Scatter plot of the measured bunching signal  in units of the second order coherence function minus 1 over time in ps for a lab measurement. A peak is visible at zero ps. The measurement appears less noisy than the measurements of the stars. }, width=\textwidth]{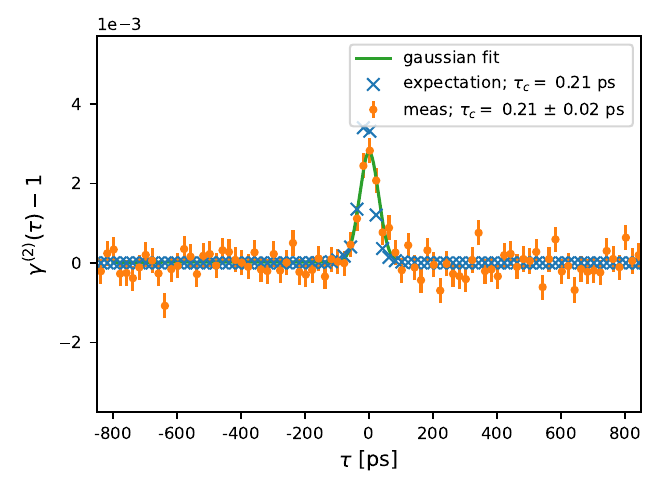}
    \caption{}
    \label{fig:lab_test}
    \end{subfigure}
    \vspace{1ex}%
\caption{(a) Timing jitter measured with the full photo detection system of 2 HPDs, 2 CFDs and the quTAG, as well as one 50:50 non-polarizing beam splitter. The delta like pulses were supplied by a frequency tripled femtosecond IR laser. At approx. $2\,$MHz count rate a FWHM timing resolution of $42.4\,$ps is observed. The measurement is rescaled to have a unit integral. (b) Bunching signal measured in the laboratory using an XBO. The measurement result fits well to the expectation, giving a measured coherence time $\tau_{\mathrm{meas, 2f-2f}} = (0.21\pm 0.02)\,$ps for an expected coherence time of $0.212\,$ps.}
\end{figure}

\FloatBarrier
In order to probe the timing resolution of our detection setup, we used laser pulses with a pulse duration much smaller than the expected timing resolution of the detection setup. We supply these pulses by single pass frequency tripling of a $1560\,$nm IR pulsed laser (Toptica FemtoFErb 1560) with a nominal pulse duration $<80\,$fs using a PPLN crystal. Since the crystal is only $0.5\,$mm long, the tripled light at $520\,$nm has a nearly identical pulse duration as the fundamental pulses. 
After frequency tripling, the intensity of the laser is strongly attenuated, in order to ensure on average less than 1 photon per laser pulse being incident on each of the two detectors. With a $100\,$ MHz laser repetition rate the detectors would otherwise be saturated quickly. 
After attenuation the light is split and directed to the HPDs by the same 50:50 non-polarizing beam splitter used in the bunching measurements.  
At approx. $2\,$MHz count rate we observed a FWHM timing resolution of $42.4\,$ps. The result of this measurement is plotted in Fig.~\ref{fig:timing_jitter}. This high resolution measurement is used to calculate the measurement expectation for the bunching measurements.
\section{Lab test results} \label{sec:lab_test_results}
\begin{figure}
\begin{subfigure}{0.5\textwidth}
	\centering
    \includegraphics[alt={Plot of the raw second order coherence function over time in ns for a laboratory test. Five red shaded peaks are visible above a noisy baseline as well as a green shaded peak.}, width=\textwidth]{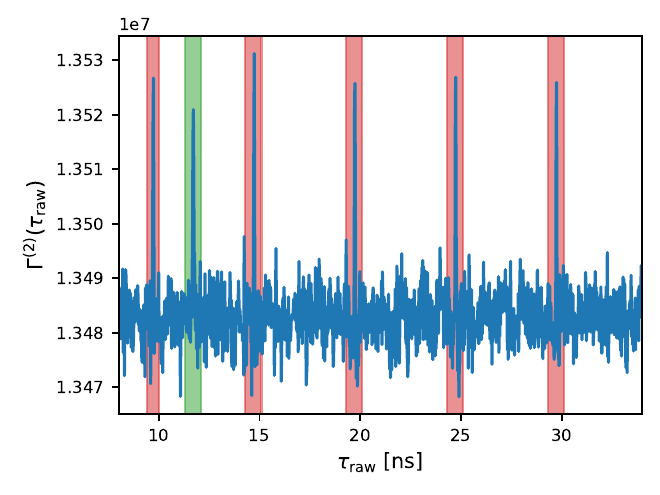}
    \caption{}
    \label{fig:lab_test_raw}
\end{subfigure}
\begin{subfigure}{0.5\textwidth}
    \centering%
    \includegraphics[alt={Root mean square error (RMSE) (top) and re-scaled residuals of the RMSE from shot noise (bottom) plotted over the number of counts per bin N. Top plot shows a square root dependence for the expectation as well as the result for a laboratory test. The curves start to differ for higher N. Bottom plot depicts the growing residuals in a two-setps like manner.}, width=\textwidth]{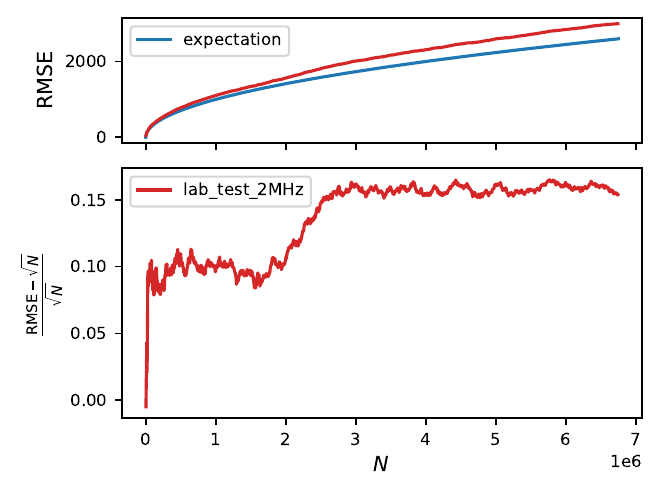}
    \caption{}
    \label{fig:RSME_residuals_lab}
\end{subfigure}
    \vspace{1ex}%
\caption{(a) Raw correlation histogram measured as a laboratory test. The area shaded in green is the area of the bunching peak, while the areas shaded in red signify the location of spikes of the quTAG's readout clock. These areas were excluded in all RMSE evaluations. (b) Root mean square error (RMSE) (top) and re-scaled residuals of the RMSE from shot noise (bottom) plotted over the number of coincidences per bin $N$. The RMSE is expected to be shot noise $\mathrm{RMSE} = \sqrt{N}$ for $N$ counts per histogram bin. The upper plot shows deviations from shot noise in a laboratory test for higher count rates. These deviations are shown in greater detail in the lower plot. Two nearly stable plateaus are visible. For smaller $N$ the residuals can be compared to the ones obtained at the telescope. However, due to the long measurement time in the laboratory test the residuals increase.}
\end{figure}

Laboratory tests were carried out using the setup detailed in Sec.~\ref{sec:2f2f}. A test at $2\,$MHz count rate per detector is discussed more closely in this section. \\
In terms of measured coherence time, the laboratory test fits our expectations extremely well, with $(0.21 \pm 0.02)\,$ps and an expected coherence time of $0.21\,$ps (cf. Fig.~\ref{fig:lab_test}). To measure a normalized second order photon correlation function, such as the one shown in Fig.~\ref{fig:lab_test}, some preprocessing is necessary. A raw correlation measurement is shown in Fig.~\ref{fig:lab_test_raw}. Note that here the second order correlation function $\Gamma^{(2)}(\tau_{\text{raw }})$ is not normalized. Besides the correlation peak (highlighted in green), some other regularly spaced peaks (highlighted in red) are visible. These peaks likely correspond to the frequency of the readout clock of the quTAG. The sections between these clock spikes are reasonably close to being shot noise limited. Thus, a $3\,$m cable delay is introduced such that the correlation peak lies inside one of these regions. The clock spikes shaded in red, as well as the region of the correlation peak shaded in green, were not used for the RMSE evaluations presented in this paper. Similarly, only the non-shaded ares of the raw correlation histogram were used in the determination of the average counts per bin, used to calculate the normalized second order correlation function.\\
Even in the laboratory, our detection electronics did not conform to shot noise perfectly. In fact, the deviations were more significant than at the telescope, finally saturating at a RMSE 15 percent larger than the shot noise expectation (cf. Fig~\ref{fig:RSME_residuals_lab}). The step in the relative RMSE at about $2 \times 10^6$ counts per bin is likely due to the long uninterrupted uptime (more than $42\,$h) of the detection system, as parameters such as the count rate or the ambient temperature did not fluctuate during the lab test.

\section{HPD quantum efficiency} \label{sec:hpd_qe}
\begin{figure}
\centering
	\includegraphics[alt={Scatter plot of the quantum efficiency over the wavelength in the range from 375 to 600 nm. The quantum efficiency slowly decreases for higher wavelengths.}, width=0.5\textwidth]{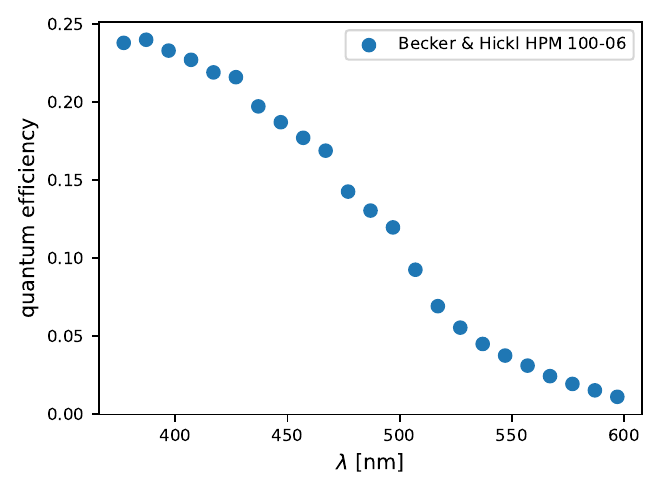}
    \caption{Quantum efficiency of a Becker \& Hickl HPM 100-06 (designated HPD in this text). At $405\,$nm a quantum efficiency of 22.7 percent is reached.}
    \label{fig:hpd_qe}
\end{figure}

The quantum efficiency of our HPDs was measured utilizing a grating based monochromator together with a XBO as a broadband light source of which the wavelengths can be selected freely. To facilitate this, the light from the XBO is collected by a condenser lens and filtered by the monochromator down to a bandwidth of $1\,$nm in the VIS depending on the width of the monochromator exit slit. The light is spatially filtered to a diameter smaller than the active area of the HPD, in order to not underestimate the quantum efficiency. \\
The quantum efficiency measurement is performed in three steps: first a calibrated photodiode together with a pico-amperemeter is used to collect a reference flux measurement at the desired wavelengths. This photodiode has an area larger than the active area of the HPD. In step two the same photodiode is used to measure the attenuation of a set of ND filters at the same wavelengths. These ND filters are necessary to attenuate the XBOs flux to a level the HPD can tolerate. In step three the HPD is used in the light beam attenuated by the ND filters calibrated in step two. For this measurement the HPD is connected directly to the quTAG without any signal conditioning by a CFD. Instead the pulse detection threshold at the quTAG is set to be just above the voltage level of HPD noise, insuring all single photon detection signals are discretized. The data necessary for the quantum efficiency measurement is acquired continuously, with the monochromators optical shutter acting as the signal to changing of the wavelength incident on the HPD. \\
Using this setup we measured quantum efficiencies slightly lower than supplied by the HPD manufacturer. An exemplary measurement result is plotted in Fig.~\ref{fig:hpd_qe}, showing a quantum efficiency of 22.7 percent at $405\,$nm, and a steep decline in quantum efficiency even at green wavelengths. This behaviour is characteristic for a Bialkali photocathode. The measured quantum efficiency is approximately 10 percent worse than specified by the manufacturer. 

\FloatBarrier
\subsection*{Disclosures}
The authors declare that there are no financial interests, commercial affiliations, or other potential conflicts of interest that could have influenced the objectivity of this research or the writing of this paper.

\subsection* {Code, Data, and Materials Availability} 
The data directly supporting the plots is available at \sloppy\url{https://doi.org/10.22000/hxRfOyGSdOgbsTJm}. The raw photon event stream can be supplied upon reasonable request from the corresponding author.

\subsection* {Acknowledgments}
We cordially thank the team of the Observatoire de la Côte d'Azur for letting us perform measurements at one of their telescopes. We also thank Nolan Matthews, Robin Kaiser, William Guerin and Mathilde Hugbart for helpful discussions on how to perform the measurement at the Omicron telescope. We acknowledge Oleg Kalekin for his support in recording the quantum efficiency measurements. We
acknowledge the financial support of the French National Research
Agency (project I2C, ANR-20-15CE31-0003).


\bibliography{report}   
\bibliographystyle{spiejour}   


\vspace{2ex}\noindent\textbf{Verena G. Leopold} is a doctoral researcher in the Qunatum Optics and Quantum Information group at Friedrich-Alexander university (FAU) Erlangen-Nuremberg. She received her BS and MS degrees in physics from FAU in 2019 and 2021, respectively. Her current research interests include stellar intensity interferometry, single photon counting, and timing hardware. She is a member of SPIE.
\\
\vspace{1ex}
\noindent Biographies and photographs of the other authors are not available.

\listoffigures
\listoftables

\end{document}